\documentclass[twocolumn,showpacs,aps,prb,superscriptaddress,floatfix]{revtex4}
\usepackage{graphicx}
\usepackage{dcolumn}
\usepackage{bm}
\usepackage{topcapt}
\usepackage{amsmath}

\usepackage{color}

\usepackage{placeins}

\begin{document}

\preprint{}

\title{Strict limit on in-plane ordered magnetic dipole moment in URu$_2$Si$_2$}

\author{K.A. Ross}
\affiliation{Institute for Quantum Matter and Department of Physics and Astronomy, Johns Hopkins University, Baltimore, Maryland 21218, USA}
\affiliation{NIST Center for Neutron Research, National Institute of Standards and Technology, Gaithersburg, Maryland 20899, USA}
%

%

\author{L.Harriger} 
\affiliation{NIST Center for Neutron Research, National Institute of Standards and Technology, Gaithersburg, Maryland 20899, USA}
%
\author{Z. Yamani} 
\affiliation{Chalk River Laboratories, National Research Council, Chalk River, Ontario K0J 1J0, Canada}

\author{W.J.L. Buyers} 
\affiliation{Chalk River Laboratories, National Research Council, Chalk River, Ontario K0J 1J0, Canada}

\author{J.D. Garrett} 
\affiliation{Brockhouse Institute for Materials Research, McMaster University, Hamilton, Ontario, L85 4M1, Canada}

\author{A.A. Menovsky} 
\affiliation{Van der Waals-Zeeman Laboratory, University of Amsterdam 1018 XE, The Netherlands}

\author{J.A. Mydosh}
\affiliation{Kamerlingh Onnes Laboratory, Leiden University, NL-2300 RA Leiden, The Netherlands}

\author{C.L. Broholm} 
\affiliation{Institute for Quantum Matter and Department of Physics and Astronomy, Johns Hopkins University, Baltimore, Maryland 21218, USA}
\affiliation{NIST Center for Neutron Research, National Institute of Standards and Technology, Gaithersburg, Maryland 20899, USA}
\affiliation{Quantum Condensed Matter Division, Oak Ridge National Laboratory, Oak Ridge, Tennessee 37831, USA}


\bibliographystyle{prsty}

%

\begin{abstract}
Neutron diffraction is used to examine the polarization of weak static antiferromagnetism in high quality single crystalline URu$_2$Si$_2$. As previously documented, elastic Bragg-like diffraction develops for temperature $T<T_{HO}= 17.5$~K at ${\bf q}=(100)$ but not at wave vector transfer ${\bf q}=(001)$. The peak width indicates correlation lengths $\xi_c=230(12)$ \AA \ and $\xi_a=240(15)$ \AA. The integrated intensity of the $T-$dependent peaks corresponds to a sample averaged $c$-oriented staggered moment of $\mu_{c}=0.022(1) \mu_B$ at $T=1.7$~K. The absence of $T-$dependent diffraction at ${\bf q}=(001)$  places a limit $\mu_{\perp}<0.0011 \mu_B$ on an $f-$ or $d-$orbital based in-plane staggered magnetic dipole moment, which is associated with multipolar orders proposed for URu$_2$Si$_2$.
\end{abstract}


\maketitle


\section{\label{sec:level1}Introduction}

Over two decades of concerted theoretical and experimental effort has so far failed to reveal the true nature of the so-called ``hidden order'' (HO) in the heavy-fermion material, URu$_2$Si$_2$ \cite{mydosh2011colloquium}.  The enigmatic HO state is signaled by a sharp specific heat anomaly at $T_{HO}$ = 17.5 K \cite{palstra1985superconducting, maple1986partially}, but the nature of the underlying order parameter is unresolved.  This state of affairs has fueled an active subfield of research, with many creative theoretical proposals, and a full battery of experimental techniques brought to bear on the problem.\cite{mydosh2011colloquium}  The eventual explanation for the HO in centered tetragonal URu$_2$Si$_2$ will need to account for numerous intriguing experimental findings, including strong fermi-surface (FS) reconstruction below T$_{HO}$ \cite{palstra1985superconducting,maple1986partially}, an antiferromagnetic phase that is reached from the HO phase by applying pressure \cite{amitsuka1999} but does not appear to affect the FS,\cite{nakashima2003haas, PhysRevLett.105.216409} $c$-axis polarized spin fluctuations which become gapped and coherent at T$_{HO}$ \cite{wiebe2007gapped}, and most recently, the identification of C4 rotational symmetry breaking of the weak basal-plane spin susceptibility. \cite{okazaki2011rotational}

\begin{figure}[!htb]  
\centering
\includegraphics[ width=\columnwidth]{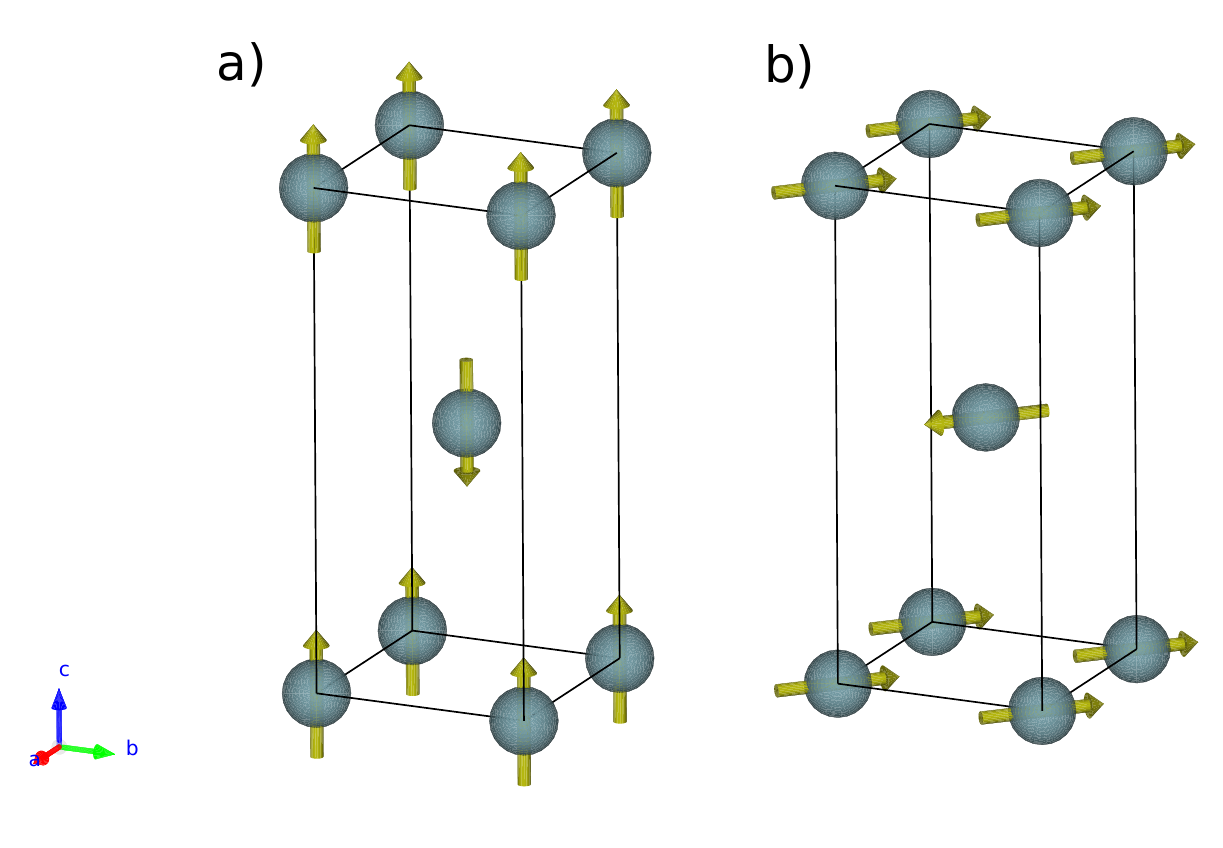}
\caption{Schematic of magnetic ordering for the a) known out of plane magnetic moments ($\mu_\parallel$) in the SMAF phase below $T_{HO}$, and b) proposed in-plane magnetic moments ($\mu_\perp$) from Refs. \onlinecite{rau2012hidden, ikeda2012emergent, Chandra2013}.}
\label{fig:moments}
\end{figure}

One of the first experimental signals which seemed to correlate with T$_{HO}$ was found using elastic neutron scattering.  The first neutron studies revealed the onset of the so-called Small Moment Antiferromagnetic (SMAF) phase below T$_{HO}$, which was characterized by a (100) ordering wave vector, and a very small ordered moment of approximately 0.03 $\mu_B$.\cite{broholm1987magnetic, mason1990neutron, isaacs1990, broholm1991magnetic}  The magnetic structure in the SMAF phase consists of dipole moments oriented along the $c$-axis, with ferromagnetically ordered layers alternating anti-ferromagnetically along the $c$-axis (Figure \ref{fig:moments} a).  The reported size of the ordered moment varies from 0.011 to 0.03 $\mu_B$/U in single crystal samples.\cite{broholm1987magnetic, mason1990neutron, isaacs1990, broholm1991magnetic, Amitsuka2007}  It is now clear that lattice strain can induce stronger sample averaged staggered magnetization possibly through inclusions of droplets of the pressure induced large moment phase,\cite{luke1994muon,matsuda2001, amato2004weak, niklowitz2010} and $^{29}$Si NMR places an upper limit of 0.0002 $\mu_B$ on a homogeneous $c$-axis oriented staggered moment that is static on the microsecond time scale (Fig. 1(a)). \cite{takagi2007no} High quality samples at ambient pressure are generally found to have a $\sim$0.01 to 0.02 $\mu_B$ staggered magnetization, with a correlation length in excess of 200 \AA \, as in the present experiment.  Whether intrinsic or extrinsic in origin, the size of the ordered moment in the SMAF phase is insufficient to explain the size of the specific heat anomaly at 17.5 K within the conventional framework for rare earth and actinide magnetism,\cite{jensen1991rare} and thus the true nature of the order parameter remains elusive.

A recent study of the magnetic susceptibility of small crystals of URu$_2$Si$_2$, using the highly sensitive torque magnetometry technique, has revealed C4 symmetry breaking in the basal plane of the tetragonal unit cell.\cite{okazaki2011rotational}  This symmetry breaking onsets at $T_{HO}$ in small crystals, indicating the formation of domains of the broken symmetry state.  Since this discovery, additional experiments have confirmed the presence of C4 symmetry breaking through NMR line width broadening \cite{kambe2013nmr} and an anisotropic cyclotron resonance signal \cite{tonegawa2012cyclotron}.  Theoretical proposals have been put forth to explain the C4 symmetry breaking, including spin nematic order \cite{fujimoto2011spin}, a spin-orbit density wave \cite{das2013imprints}, and a modulated spin liquid phase.\cite{pepin2011modulated} These proposals show how spin rotation symmetry could be broken while respecting time reversal symmetry, and hence avoiding the formation of a magnetic dipole moment in the basal plane.  However, another possibility is also evident: small in-plane ordered dipole moments which have so far escaped detection.  

\begin{figure}[!htb]  
\centering
\includegraphics[ width=\columnwidth]{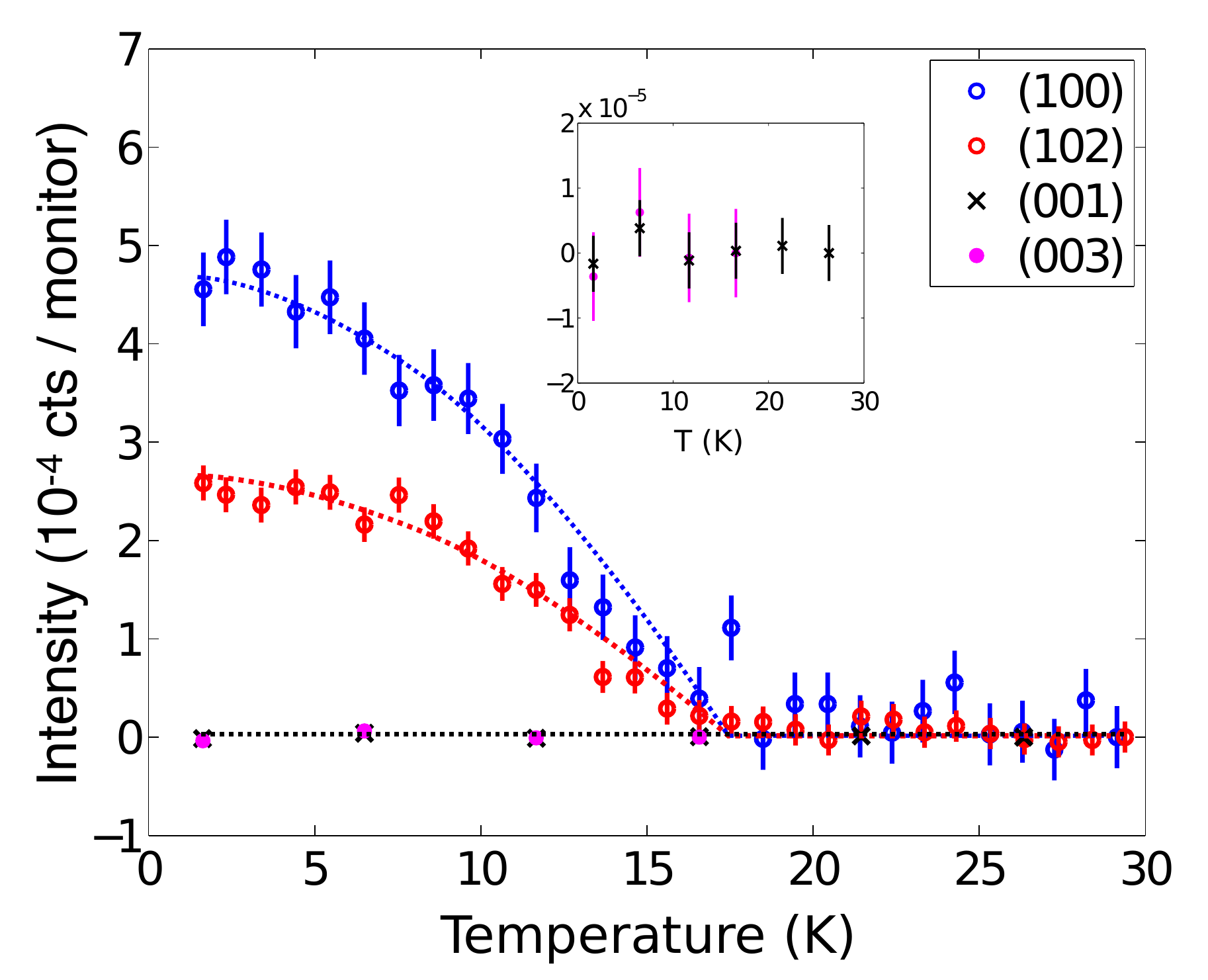}
\caption{ Temperature dependence of the peak intensity at several candidate magnetic Bragg peaks.  The inset shows the temperature dependence of the peaks which would correspond to in-plane magnetic ordering, on a finer intensity scale.  Dotted lines are guides to the eye.  The monitor count rate was 6100 counts/s. Errorbars represent one standard deviation.}
\label{fig:tdep}
\end{figure}

Though a small ordered in-plane dipole moment would not by itself account for the change of entropy at the hidden order transition any more than the $c$-oriented SMAF, its presence or absence is critical to understanding C4 symmetry breaking.  A putative transverse staggered magnetization was recently considered in three separate theoretical works.  Rau \emph{et al}\cite{rau2012hidden} proposed a spin density wave (SDW) involving 5$f$ crystal field doublets.  The (undetermined) details of the crystal field wavefunctions dictate the size of the in-plane moment induced by such a SDW, which could be vanishingly small.  Nevertheless, a small moment of unspecified magnitude pointing along [110] is expected in this case \cite{rau2012hidden}.  A second theoretical prediction of a small in-plane moment comes from Chandra \emph{et al}, who propose the order parameter characterizing the HO phase is hybridization between conduction electrons and local Ising-like 5$f^2$ wavefunctions, the combination of which produce an object that breaks both single and double time-reversal symmetry, a so-called ``hastatic'' order which entails a small in-plane moment in the hidden order phase. \cite{Chandra2013}  Chandra \emph{et al} place a theoretical upper limit of $\mu_\perp$ = 0.015 $\mu_B$ on the size of the in-plane ordered moment.  A third paper examines the complete set of multipole correlations allowed in URu$_2$Si$_2$ \cite{ikeda2012emergent}. Employing density functional theory to establish a multi band Anderson Hamiltonian and augmented RPA theory to account for interactions, a rank-5 multipole (dotriacontapole) order with `nematic' E- symmetry was found to be critical to a low temperature condensed phase. While no estimate of magnitude is provided, a staggered pseudospin moment along the [110] direction is concomitant to this order. Although the dotriacontapole order was initially supported by the interpretation of high magnetic field neutron diffraction data, \cite{ressouche2012hidden} this conclusion has recently been discounted based on a space group analysis.\cite{khalyavin2014symmetry}  Furthermore, a recent photoemission study of the FS in URu$_2$Si$_2$ \cite{meng2013imaging} did not observe features associated with dotriacontapole ordering in Ref. \onlinecite{ikeda2012emergent}.  The photoemission experiment provided evidence for itinerant 5$f$ electrons so that theories relating to (localized) crystal field wave functions such as Refs. \onlinecite{rau2012hidden} to \onlinecite{Chandra2013} may be called into question. Nevertheless, even in the absence of a fully consistent theory, in light of the C4 symmetry breaking magnetic susceptibility data, it is of utmost importance to establish whether or not a small dipole moment is formed in the basal plane of URu$_2$Si$_2$. 

We therefore performed an elastic neutron scattering experiment to search for a small ordered in-plane dipole moment in a single crystal of URu$_2$Si$_2$.  Our measurement puts an upper limit on the possible size of any such moment of $\mu_\perp$ $<$ 0.0011 $\mu_B$ assuming a magnetic form factor appropriate to the 5$f^{3}$ electron configuration of U$^{3+}$.  Because the measurement was carried out as close as possible (${\bf q}$=(001)) to the origin of reciprocal space, the limit is relatively insensitive to the electronic orbital associated with a putative staggered magnetization.

\section{\label{sec:method}Experimental Method}

\begin{figure*}  
\centering
\includegraphics[ width=1.5\columnwidth]{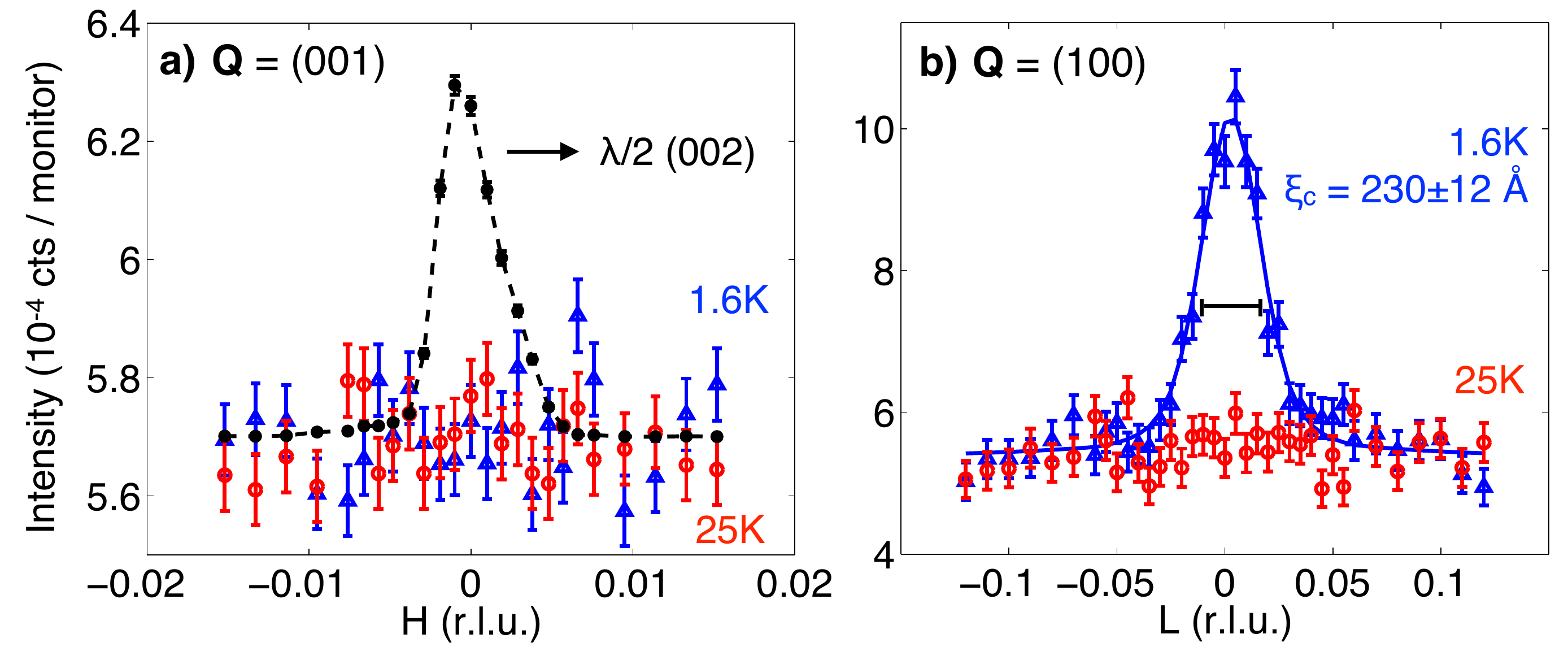}
\caption{ Transverse ${\bf q}$ scans at two positions, a) (001) and b) (100), which would correspond to in-plane and out-of-plane magnetic ordering, respectively.  Only the (100) position shows a magnetic Bragg peak upon cooling through $T_{HO}$ = 17.5 K.  A transverse scan performed at the same position as (001) but without the higher-order filters in the beam (i.e., intentionally including $\lambda/2$ contamination from the (002) nuclear Bragg peak) indicates the correct position of the scan in reciprocal space (black dashed line in a), scaled down by a factor of 7200.  The horizontal bar in b) represents the calculated full width at half maximum of the instrumental resolution (see Fig. \ref{fig:fwhm}).  The monitor count rate was 6100 counts/s. Errorbars represent one standard deviation.}
\label{fig:transverse}
\end{figure*}

A large single crystal of URu$_2$Si$_2$ was grown by the Czochralski method using a Tri-Arc furnace at McMaster University.  The crystal was oriented in the $(H0L)$ plane and mounted in a helium flow cryostat on the SPINS triple-axis spectrometer at the NIST Center for Neutron Research.   The instrument was configured for elastic scattering with $E_{i}$ = $E_{f}$ = 4.7 meV, using a vertically focussed PG(002) monochromator and a flat PG(002) analyzer. The Full Width at Half Maximum energy resolution for this configuration is 0.27 meV. Cooled beryllium filters were used before and after the sample to eliminate second order ($\lambda/2$) contamination.   SPINS has a nickel supermirror guide before the monochromator, giving an energy-dependent incident beam divergence of 1 degree \AA$^{-1}$/$k_i$, where $k_i$ is the incident wavenumber of the neutrons.  Soller collimators (80' before and after the sample) were in place to control the in-plane beam divergence. 

The magnetic neutron scattering cross section is sensitive to the relative orientation of the scattering wave-vector, ${\bf q} = {\bf k_{i}} - {\bf k_{f}}$, to the direction of the magnetic moment, ${\bf M}$.  Specifically, the cross section is maximal for ${\bf q} \perp {\bf M}$ and exactly zero for ${\bf q} \parallel {\bf M}$.   Thus, for a magnetic structure with moments oriented along the $c$-axis (Fig. \ref{fig:moments}(a)) there is no magnetic contribution to Bragg scattering at $(00L)$-type positions.  In contrast, for in-plane ordered moment, such as that proposed in Refs.  \onlinecite{rau2012hidden}, \onlinecite{ikeda2012emergent}, and \onlinecite{Chandra2013} and shown in Fig.  \ref{fig:moments}(b), the polarization factor is largest for $(00L)$-type positions.  With our experimental configuration, we can access $(H0L)$ reflections and differentiate between transverse and  in-plane ordered moments by focusing on  $(H00)$- and $(00L)$- type reflections, respectively.

The space group for URu$_2$Si$_2$ is centered tetragonal. The conventional tetragonal unit cell, which contains two formula units, has lattice parameters $a=b=4.129$ and $c = 9.573 $\AA \ at T=293 K and the general selection rule for reflections with Miller indices $(hkl)$ in the corresponding tetragonal reciprocal lattice is $h+k+l=2n$. The magnetic structures proposed (in-plane moment, Fig. \ref{fig:moments} b)) and measured ($c$-axis SMAF phase, Fig. \ref{fig:moments} a)), however, break the centering translational symmetry and produce peaks at the nuclear-forbidden positions.  This provides an ideal scenario to search for small magnetic signals on a low background at $(hkl)$ where $h+k+l=2n+1$, while differentiating between in-plane and transverse ordered moments through the polarization factor.

A note on our choice of incident energy is appropriate.  The use of cold neutrons with low incident energy allows us to analyze the elastically scattered neutrons with an energy resolution of ~0.27 meV, thereby excluding inelastic processes which could contribute to overall background and obscure a small signal.  We chose an incident energy of 4.7 meV to avoid the effects of multiple-scattering.  Possibilities for multiple-scattering, i.e. where some proportion of the detected neutrons have undergone more than one scattering process before being detected, depend on the incident energy and details of the Bragg reflection geometry.  The effect can be strong for large samples such as the one studied here to achieve sensitivity to small moment magnetic ordering.  Considering the lattice parameters for URu$_2$Si$_2$ and our use of the $(H0L)$ scattering plane, we find there are no multiple Bragg scattering processes for the (001) or (003) positions and neutron energies from 3.34 to 4.71 meV (Appendix \ref{sec:multscatt}).


\section{\label{sec:level1}Results and Discussion}

First, we looked for the SMAF ordered moment polarized along the $c$-axis.  Figure \ref{fig:tdep} shows the temperature dependence of the (100) and (102) peak intensities. Both are allowed magnetic Bragg peaks for this type of ordering (Figure \ref{fig:moments} a)).   The (100) and (102) reflections were collected at 3 minutes and 10 minutes per point, respectively, and the intensity is normalized to a monitor in the incident beam with a count rate of 6100 counts/s.  Both reflections show a temperature dependence consistent with a transition near 17.5 K, the hidden order transition temperature.  In contrast, the intensities at (001) and (003), which are sensitive to static in-plane order, show no increase above background.  For this temperature scan, the intensity at the (001) position was measured for 2.8 hours per point, while (003) was measured for 55 min per point.    Transverse scans at (001) and (100) are shown in Figure \ref{fig:transverse}, at two temperatures, 1.6 K and 25 K.  The (100) peak disappears above the transition.  The (001) position does not show a peak at either temperature.  To prove the trajectory probed actually passed through the (001) location in reciprocal space we also show data collected without the beryllium filters in the beam (black filled symbols).  Under those conditions there is a component of $\lambda/2$ neutrons that diffracts from the (002) nuclear Bragg peak.  It appears exactly where a putative (001) magnetic peak would occur for $\lambda$ neutrons.  There is no sign of a rod of scattering parallel to $c$ at (001) which would manifest as a peak in the transverse scan at both temperatures.  The lack of rods is an indication of reduced stacking faults in this crystal, which is consistent with the relatively small SMAF moment ($\mu_\parallel \simeq 0.02 \mu_B$). \cite{broholm1987magnetic, isaacs1990, Amitsuka2007} Fig. 3(b) shows transverse scans through the (100) position. From a flat background 
in the PM phase (T=25 K) a peak develops in the HO phase (T=1.6 K). This peak is not resolution limited (horizontal bar and Fig. 5(a)-(b)). A fit of perpendicular scans through (100) to a Lorentzian convoluted by a Gaussian leads to correlation lengths of $\xi_c$=230(12) \AA \ and $\xi_a$=240(15) \AA \ along the $c$ and $a$ directions respectively. Previously reported correlation lengths for URu$_2$Si$_2$ range from 200 \AA \cite{isaacs1990} to 400 \AA \cite{broholm1987magnetic} for the $a$-axis and  100 \AA \cite{broholm1991magnetic} to 450 \AA \cite{isaacs1990} for the $c$-axis.

To determine the size of the out-of-plane magnetic moment $\mu_\parallel$ and place limits on the in-plane moment $\mu_\perp$, we normalize to the (101) nuclear reflection, which has the weakest structure factor of any reflection from the URu$_2$Si$_2$ crystal structure, and therefore is least affected by extinction.  Details of the normalization are given in Appendix \ref{appen}.

\begin{figure}  
\centering
\includegraphics[ width=\columnwidth]{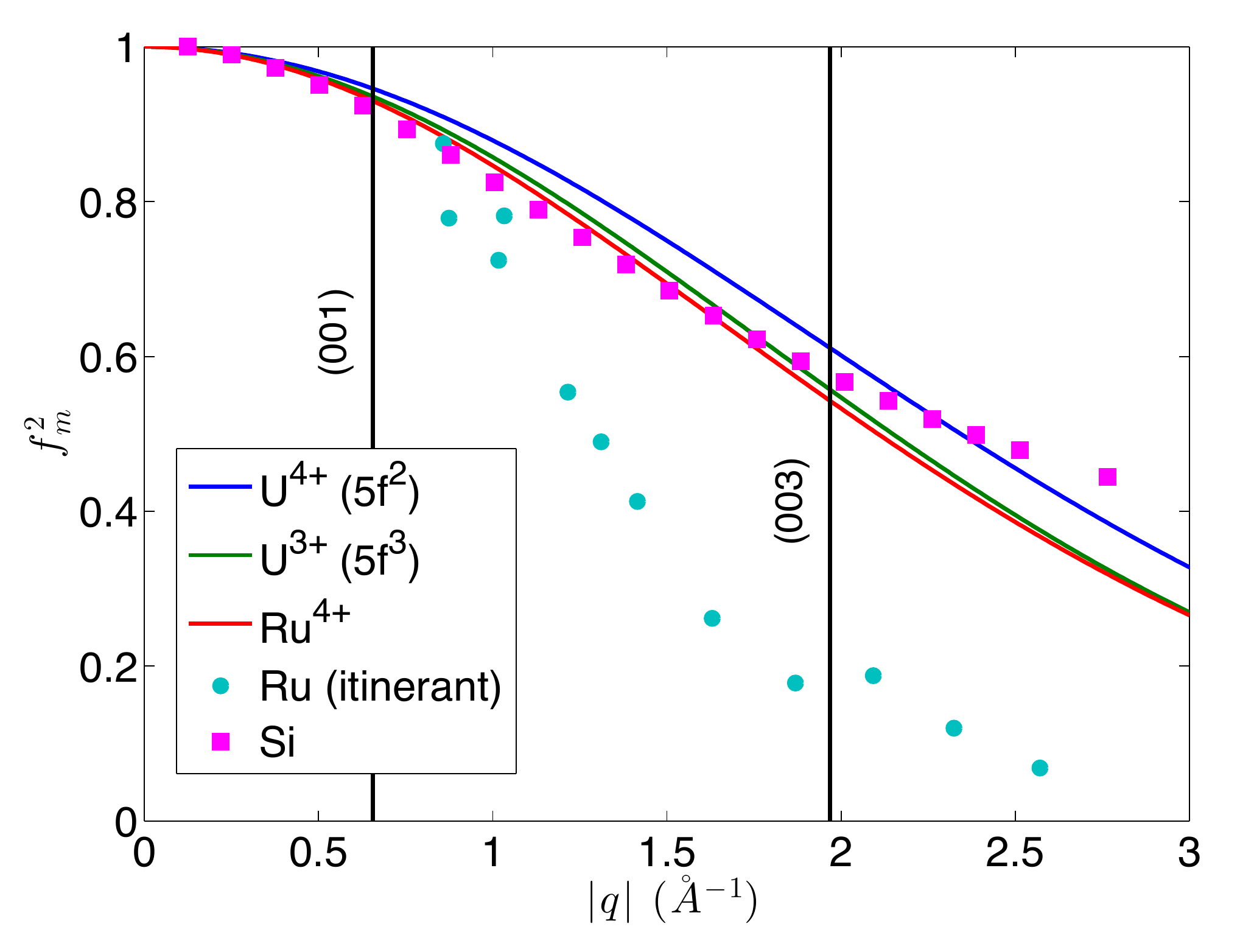}
\caption{ Square of the magnetic form factors for several types of spin densities as a function of scattering wave-vector, $|{\boldsymbol q}|$.  The first three curves correspond to localized ionic states of uranium (U$^{4+}$ (5$f^2$) and U$^{3+}$ (5$f^3$)) and ruthenium (Ru$^{4+}$), and are described by Equation \ref{eqn:formfactor}.  We also roughly approximate the possible spin density that could be associated with Ru- or Si-derived conduction bands, for example as might be appropriate for conduction electrons in the hastatic order theory \cite{Chandra2013}.  The data for Si are those of the x-ray atomic scattering factor squared, which measures the Fourier transform of the total electron density. \cite{brown2006magnetic} }
\label{fig:ffact}
\end{figure}

Entering into the magnetic cross section is the magnetic form factor, $f_M(q)$, which is the Fourier transform of the unpaired spin density and depends on the spatial distribution of spin density.  In the ``hastatic order'' scenario,  $\mu_\perp$  contains both conduction electron and 5$f^3$ components, and presumably has a more extended spin density.  Figure \ref{fig:ffact} shows the magnetic form factor for various spin density distributions. Two possibilities for a localized U moment are shown, namely the 5$f^2$ and 5$f^3$ electronic configuration along with some other possibilities that relate to moments forming from bands derived from Ru or Si orbitals. We also reproduce the measured magnetic form factor squared from itinerant magnetic excitations associated with Ru in Sr$_2$RuO$_4$ \cite{servant2002magnetic}.  For Si, we show the square of the normalized x-ray atomic scattering factor, tabulated in Ref. \onlinecite{brown2006magnetic}, which measures the Fourier transform of the total local electron density.  These latter approximations are only intended to show plausible variations in magnetic form factor around the (001) and (003) positions, and not to represent realistic calculations for spin density in URu$_2$Si$_2$.  We note that because we are basing the size of the upper limit for $\mu_\perp$ solely on a measurement at the (001) position, which has a very small wave-vector ($|{\bf q}|$ = 0.65 \AA$^{-1}$), sensitivity to the choice of form factor is small, approximately 1 \%.  Upper limits that are based on the (003) position with $|{\bf q}|$ = 1.95 \AA$^{-1}$  are affected much more severely.   Thus, though our limit is consistent with that quoted by Das \emph{et al},\cite{das2013absence} and is stricter than that quoted by Metoki \emph{et al}, \cite{metoki2013neutron} those experiments are essentially only sensitive to local 5$f$ electron magnetism, since they are based in part\cite{das2013absence} or entirely\cite{metoki2013neutron} on data from (003).


Table \ref{tab:summary} summarizes the measured intensities of the relevant magnetic and nuclear peaks, as well as the derived magnetic moments.   We use form factors based on the U 5$f^2$ and 5$f^3$ electronic configurations to determine $\mu_\parallel$ and $\mu_\perp$, respectively (5$f^3$ was chosen for $\mu_\perp$ since it produces a more relaxed limit).

 \begin{table}[htbp]
    \centering
    \begin{tabular}{@{} lllll @{}} 
       \toprule
       Reflection    & T (K) & $F_{meas}^2$ (barns) & $F_{calc}^2$ (barns) & $\mu$ ($\mu_B$)\\
\hline
       (101)      & 26 & 0.29(6) &  0.29 & -- \\
       (100)     & 1.6  & 10.0(8)$\times10^{-5}$ & -- & 0.022(1)  \\
       (102)       & 1.6  & 6.7(5)$\times10^{-5}$ & -- & 0.022(1) \\
       (001)   & 1.6   &  $<$ 3.0(2)$\times10^{-7}$  & -- & $<$ 0.0011  \\
       \hline
    \end{tabular}
    \caption{Summary of measured and calculated results for various magnetic and nuclear Bragg peaks.  The moment size listed for the (001) reflection is an upper bound on any in-plane magnetic moment.}
    \label{tab:summary}
 \end{table}

\section{\label{sec:level1}Conclusions}

Our search for an in-plane ordered dipole moment in a high quality single crystal of URu$_2$Si$_2$ places an upper limit of $\mu_\perp \leq 0.0011 \mu_B$ on the size of any such moment with a correlation length in excess of 200 \AA.  Our measurement cannot rule out ordered in-plane moments smaller than this or with a substantially shorter correlation length.  We note that it is still within the bounds placed on the dipole moment produced in an ordered \emph{dotriacontapole} model.\cite{ikeda2012emergent} The assumption of a localized 5$f^3$ character of the magnetic form factor may not be suitable for theories such as the hastatic order, where the in-plane moment arises in part from conduction electrons with a more extended spin density distribution.  In that case, the size of the magnetic moment consistent with our measurements (and others obtained from neutron scattering \cite{das2013absence,metoki2013neutron}) could be larger than the value quoted here.  With a specific form factor it would be possible to calculate a new limit based on the magnetic scattering structure factor listed in Table \ref{tab:summary}.

Our experiment thus places strict and well-defined constraints on theories with in-plane magnetic dipole moments, such as the theories of hastatic order \cite{Chandra2013} and rank-5 superspin density wave\cite{rau2012hidden}.   This may shift the focus to theories where C4 rotational symmetry is broken while retaining time-reversal symmetry.

The authors acknowledge illuminating discussions with P. Chandra, P. Coleman and R. Flint.   K.A.R. acknowledges the helpful logistical assistance of J.W. Lynn.   Work at IQM was supported by the US Department of Energy, office of Basic Energy Sciences, Division of Material Sciences and Engineering under grant DE-FG02-08ER46544.  This work utilized facilities supported in part by the National Science Foundation under Agreement No. DMR-0944772.

\appendix

\section{Normalized Neutron Diffraction}
\label{appen}
The elastic magnetic neutron scattering cross section at a reciprocal lattice vector $\boldsymbol \tau$ associated with a periodic magnetic structure is, \cite{squires2012introduction}

\begin{equation}
\bigg(\frac{d\sigma}{d\Omega}\bigg)_{M}({\bf q}) = (\gamma r_0)^2 \frac{N (2\pi)^3}{v_{0}} \sum_{{\boldsymbol \tau}} \delta({\bf q} - {\boldsymbol \tau}) |\hat{\bf q} \times {\bf F_M} \times \hat{\bf q}|^2,
\label{eqn:mag}
\end{equation}
 where $\gamma = 1.193$ is the magnetic dipole moment of the neutron in units of the nuclear Bohr magneton, $r_0 = 2.818\times10^{-12}$ cm is the classical electron radius, and ${\bf F}_M$ is the magnetic structure factor, which, for collinear moments, is given by
                                      
\begin{equation}
{\bf F}_M({\boldsymbol \tau}) = \frac{1}{2} g_L \langle{\bf J}\rangle f_M({\boldsymbol \tau}) \sum_{\bf d}\sigma_{\bf d} \exp{(i {\boldsymbol \tau} \cdot {\bf d})} \exp(-W_{\bf d}),
\end{equation}           
where $g_L$ is the Land\'e g-factor, $\langle{\bf J}\rangle$ is the expectation value of the angular momentum operator, $f_M(\boldsymbol{\tau})$ is the magnetic form factor (Eqn. \ref{eqn:formfactor}), $\exp(-W_{\bf d})$ is the Debye-Waller factor, which accounts for atomic thermal motion and which is indistinguishable from unity at temperatures and wave vectors of interest, $\sigma_{{\bf d}} = \pm 1$, and the sum is over the $\bf{d}$ atom basis.  In the case of URu$_2$Si$_2$, the  basis consists of two ions at ${\bf d_1} = (0,0,0)$ and ${\bf d_2} = (1/2,1/2,1/2)$.  The magnetic structure factor reduces to,
\[
\begin{array}{lll}
	{\bf F}_M(hkl)={\boldsymbol \mu} \cdot f_M(hkl) & , & \text{ for $h+k+l$ = odd} \\
	{\bf F}_M(hkl) = 0 & , & \text{ for $h+k+l$ = even}
\end{array}
\]

 For the SMAF structure ${\boldsymbol \mu \parallel \bf c}$, producing non-zero intensity at (100), (300) etc.  For the proposed in-plane moments, ${\boldsymbol \mu \perp \bf c}$, and intensity should arise at (001), (003), etc.  Details of the magnetic form factors, $f_M$, are given in Appendix \ref{sec:formfactor}.

\begin{figure*}[!htb]  
\centering
\includegraphics[width=1.5\columnwidth]{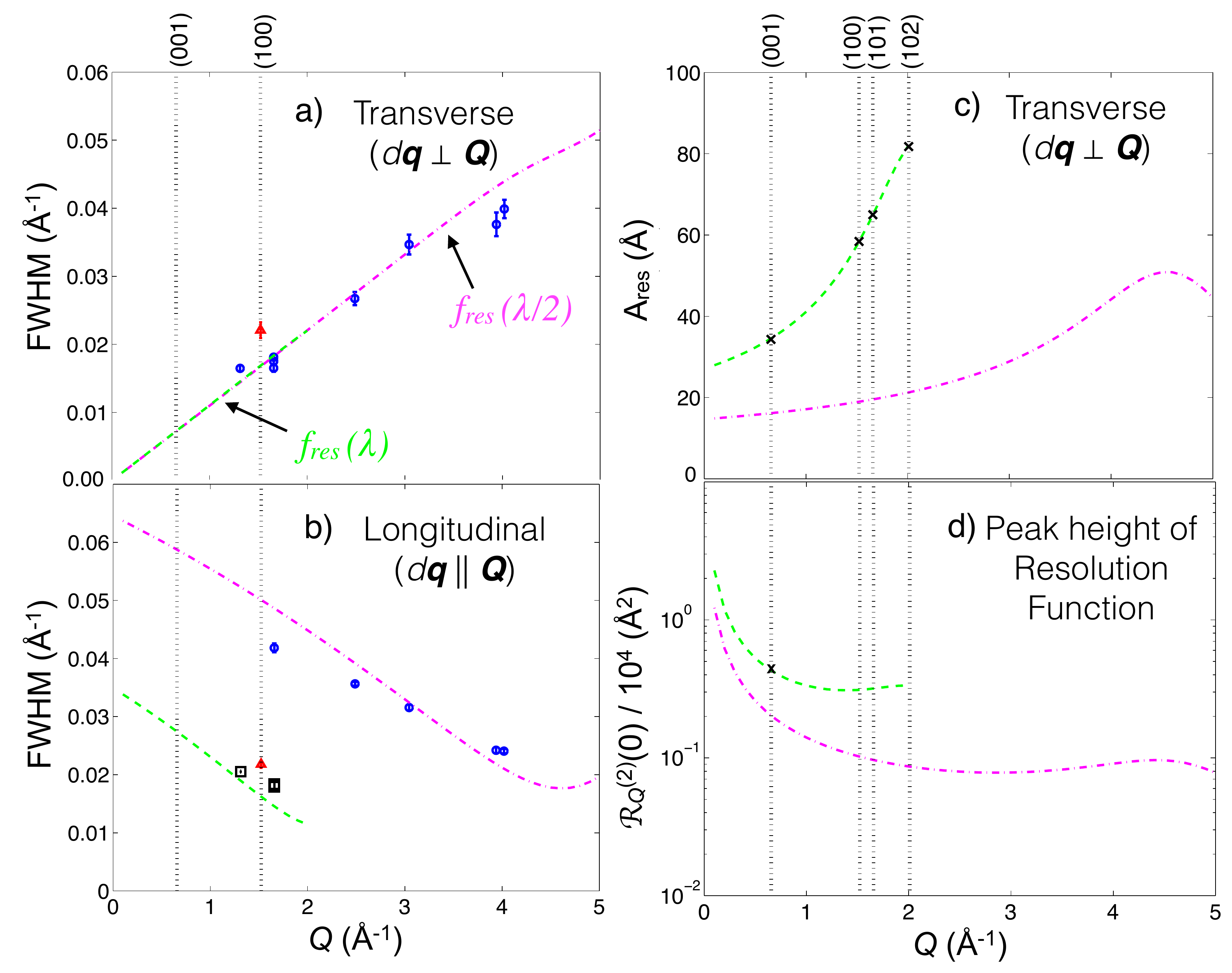}
\caption{ a) and b) Calculated and measured FWHM of Bragg peaks as a function of scattering wavenumber $q$, for transverse and longitudinal scans.  The blue circles ($\lambda$/2) and black squares ($\lambda$) show the measured FWHM for nuclear peaks, and the red triangle shows the measured FWHM for the (100) magnetic peak (errorbars represent half of the 95\% confidence interval resulting from fits to a gaussian lineshape).   The calculated FWHM ($f_{res}$) are based on resolution function calculations for the following instrumental configuration: incident beam divergence of 60'/$k_i$,  PG(002) vertically focussed monochromator, 80' collimators before and after the sample, PG(002) flat analyzer, and 33' sample mosaic.  For $Q <$ 2 \AA$^{-1}$ the FWHM are measured and calculated (green dashed line) for incident neutrons with $k_i$=2$\pi$/$\lambda_{PG(002)}$ =  1.8733 \AA$^{-1}$.  For $Q >$ 2 \AA$^{-1}$, the FWHM are shown (pink dot-dashed line) for $k_i$=2$\pi$/$(\lambda_{PG(002)}/2)$ =  3.7465 \AA$^{-1}$.  c) Calculated areas of the resolution function ($A_{res}$) for transverse scans as a function of $Q$, for the instrumental configuration above (green dashed line is for $\lambda_{PG(002)}$, pink dot-dashed line is for $\lambda_{PG(002)}$/2). d) Peak height of the normalized two dimensional resolution function (Eqn. \ref{R2D}). The black crosses show the values used for determining the measured cross sections at the specified zone centers.}
\label{fig:fwhm}
\end{figure*}

\begin{table*}
   \centering
   \begin{tabular}{|c|c|c|c|c|c| } 
     \hline
      Refl.    & $|Q|$ &  $I(Q)$ &  \multicolumn{2}{c|}{$\mu$ $(\mu_B)$ $^{c}$}& moment  \\  \cline{4-5}
     & (\AA$^{-1}$)  & ($10^{-7}$ counts/monitor)& $|f_{M}|^2_{5f^2}$ & $|f_{M}|^2_{5f^3}$ & direction\\
     \hline
      (100)	& 	1.5217 	& 		2.8(1)$^a$	  &	0.0215(8)  & 0.0222(9) & $\parallel c$\\
      (102)   & 	2.0097     	& 		2.0(1)$^a$   &	0.0201(9) &  0.0210(9) &$\parallel c$ \\ 
      (001)   &	0.6563	&		 0.0085$^b$ &   $<$ 0.00105(3) & $<$ 0.00106(3) &  $\perp c$ \\
      (101)   &	1.6572 	&  		8200(500)$^a$  & -- & -- & --\\    
      \hline         
     
   \end{tabular}
   \caption{ Data obtained from magnetic reflections and the (101) nuclear reflection, leading to the determination of $\mu_{\perp max}$ and $\mu_{\parallel}$.  \\
 $^{a} I(Q)$ is determined using the integrated area of a transverse q-scan across the peak of interest ($A_{meas}$), via $I(Q) = A_{meas} / A_{res}$ where $A_{res} = \int_{\boldsymbol \tau} d{\bf q}^\prime{\cal R}_{\boldsymbol \tau}^{(2)}({\bf q}^\prime-\boldsymbol{\tau})$ is the area of a transverse scan through the two-dimensional resolution function (normalization defined in Eqn. \ref{R2D}), plotted in Fig. \ref{fig:fwhm}c) as a function of $|Q|$.  \\
 $^{b}I(Q)$ is determined from the maximum peak height, $P_{max}$, via $I(Q) = P_{max}/{\cal R}_{\bf Q}^{(2)}({\bf 0})$.  \\
 $^{c}$  Moments are listed for two choices for the magnetic form factor, as shown in Fig. \ref{fig:ffact}. }
   \label{tab:magneticdata}
\end{table*}

The coherent elastic nuclear scattering cross section is given by,

\begin{equation}
\bigg(\frac{d\sigma}{d\Omega}\bigg)_{N}({\bf q}) = \frac{N (2\pi)^3}{v_{0}} \sum_{{\boldsymbol \tau}} \delta({\bf q} - {\boldsymbol \tau})|F_N({\bf q})|^2,
\label{eqn:nuclear}
\end{equation}
where $F_N$  is the nuclear structure factor, given by
\begin{equation}
F_N({\bf q}) = \sum_{{\bf d}} b_{{\bf d}} \exp{(i {\bf q}\cdot {\bf d})\exp(-W_{\bf d})},
\end{equation}
where $b_{{\bf d}}$ is the coherent scattering length for the atom which is located at the basis vector ${\bf d}$, and $\exp(-W_{\bf d})$ is the Debye-Waller factor. \\

Denote a three-dimensional resolution function for elastic diffraction as ${\cal R}_{\bf Q}^{(3)}({\bf q-Q})$  with the same normalization condition as the Dirac $\delta$-function: 
 \begin{eqnarray}
\int d^3{\bf q} {\cal R}_{\bf Q}^{(3)}({\bf q})=1.
\end{eqnarray}
The monitor-normalized intensity near a nuclear Bragg peak, $\boldsymbol \tau$, takes the following form: 
\begin{equation}
{\cal I}_N({\bf q})={\cal C} \frac{N (2\pi)^3}{v_{N}}  |F_N(\boldsymbol{\tau})|^2 {\cal R}_{\boldsymbol \tau}^{(3)}({\bf q-\boldsymbol\tau}).
\end{equation}
Here ${\cal C}$ is a suitably dimensioned pre-factor, which absorbs all sensitivity related factors characterizing a given instrumental configuration. 

Containing the identical pre-factor, the  corresponding expression for magnetic scattering under the same experimental conditions reads
\begin{equation}
{\cal I}_M({\bf q})={\cal C} (\gamma r_0)^2\frac{N (2\pi)^3}{v_0}  |{\bf F}_{M\perp}(\boldsymbol{\tau})|^2 {\cal R}_{\boldsymbol \tau}^{(3)}({\bf q- \boldsymbol \tau}).
\end{equation}
Here we have defined the transverse projection of the magnetic vector structure factor: 
\begin{equation}
{\bf F}_{M\perp}(\boldsymbol{\tau})=\hat{\boldsymbol\tau}\times {\bf F}_M(\boldsymbol{\tau})\times \hat{\boldsymbol\tau}
\end{equation}
If, as for $\rm URu_2Si_2$, the chemical structure and therefore the nuclear scattering cross section is well known, the dimensionless squared magnetic structure factor can be determined through ratios of peak intensities as follows:
\begin{equation}
\label{peak}
|{\bf F}_{M\perp}(\boldsymbol{\tau}_M)|^2=\frac{|F_N(\boldsymbol{\tau}_N)|^2}{(\gamma r_0)^2} \frac{{\cal I}_M({\boldsymbol{\tau}_M})}{{\cal I}_N({\boldsymbol{ \tau}_N})}\frac{{\cal R}_{\bf \tau_N}^{(2)}({\bf 0})}{{\cal R}_{\bf \tau_M}^{(2)}({\bf 0})}.
\end{equation} 
Note that by using the fact that the resolution perpendicular to the scattering plane is independent of the scattering angle, we have split off that part of the resolution function so as to focus on the two-dimensional in-plane elastic resolution function with the following normalization condition: 
 \begin{eqnarray}
 \label{R2D}
\int d^2{\bf q} {\cal R}_{\bf Q}^{(2)}({\bf q})=1.
\end{eqnarray}  
The magnetic structure factor can also be obtained from a ratio of integrated intensities
\begin{eqnarray}
\label{integrated}
|{\bf F}_{M\perp}(\boldsymbol{\tau}_M)|^2&=&\frac{|F_N(\boldsymbol{\tau}_N)|^2}{(\gamma r_0)^2} \frac{\int_{\boldsymbol \tau_M} d{\bf q}^\prime {\cal I}_M({\bf q}^\prime)}{\int_{\boldsymbol \tau_N} d{\bf q}^\prime{\cal I}_N({\bf q}^\prime)}\nonumber \times \\
&&\frac{\int_{\boldsymbol \tau_N} d{\bf q}^\prime{\cal R}_{\boldsymbol \tau_N}^{(2)}({\bf q}^\prime-\boldsymbol{\tau}_N)}{\int_{\boldsymbol \tau_M} d{\bf q}^\prime{\cal R}_{\boldsymbol\tau_M}^{(2)}({\bf q}^\prime- {\boldsymbol \tau}_M)}
\end{eqnarray} 
Here it is understood that the path of integration is matched for the measured integrated intensity and the corresponding integral over the resolution function. In Equations~\ref{peak} and \ref{integrated} the measured intensity ratio is balanced by the corresponding ratio for the resolution function. If $\boldsymbol \tau_N\approx \boldsymbol \tau_M$ the latter ratio will not deviate significantly from one. 

The resolution function can readily be calculated based on the instrumental configuration.\cite{cooper1967resolution} Fig.~5 shows the wave vector dependence of various aspects of the resolution function. By comparing the experimentally measured width of nuclear Bragg peaks with the calculated widths, Figs.~5(a) and (b) indicate that the calculation based on the known beam divergences and the instrument geometry accurately reflects the instrumentation configuration. Fig.~5(c) shows the wave vector dependence of the transverse (rocking) integral through the resolution function for use in Eq.~\ref{integrated} and Fig.~5(d) depicts the q-dependence of the peak value of the normalized two dimensional elastic resolution function for use in Eq.~\ref{peak}.

Table II shows the values for the $c$-oriented staggered magnetization calculated from the (100) and (102) magnetic peak intensities using Eq.~\ref{integrated}. \\

\begin{figure}[htbp]  
\centering
\includegraphics[width=8cm]{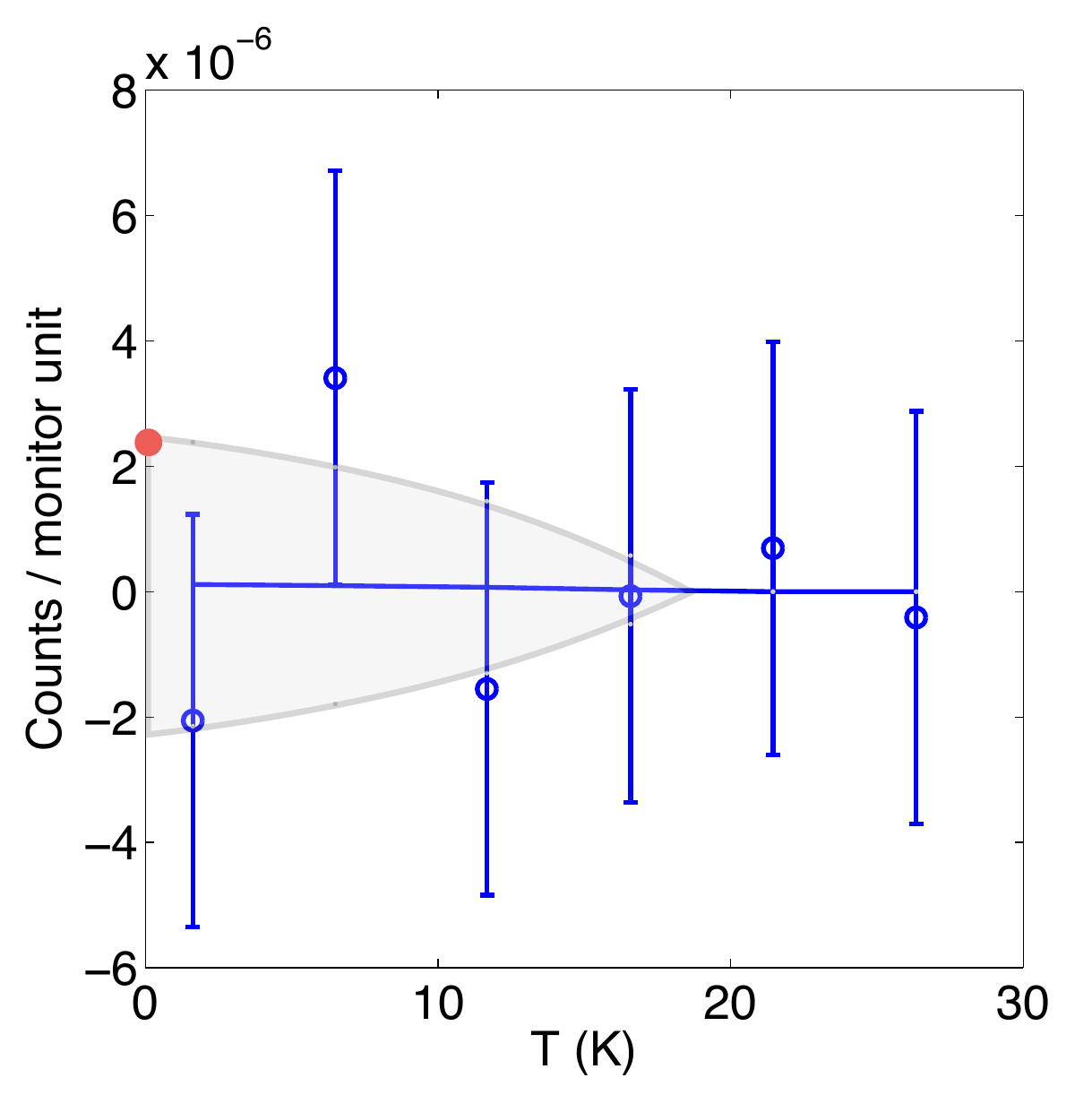}
\caption{ Blue circles: the temperature dependence of intensity at (001), subtracting the mean value, with a fit to an order parameter curve with fixed $\beta = 0.5$, $T_c=17.5 K$ (blue solid line).  The range corresponding to one standard deviation in $I_0$ is shown as a gray shaded region.  The maximum possible $I_{0_{max}} = 2.50\times10^{-6}$ counts/monitor (red circle), is used to determine the upper limit on $\mu_\perp$.}
\label{fig:orderparam}
\end{figure}

%
%
%

To determine the upper bound on $\mu_\perp$, we fit the intensity vs. temperature data at (001) to the following functional form,

\[
\begin{array}{llr}
I(T) = I_0 (T_c - T)^{2\beta} &    ,    &T<T_c \\
I(T) = 0&  ,    & T>T_c 
\end{array}
\]

This form represents an order parameter curve, as could be expected for a 2nd order phase transition.   We fix $\beta = 0.5$ (in the absence of further information about the universality class of the transition) and $T_c = 17.5$ K and allow $I_0$ to vary.  The fitted value is $I_0 \equiv (0.1 \pm  2.5) \times 10^{-6}$, and we take the maximum possible value of $I_0$ consistent with this fit to determine the upper limit on the in plane moment, $\mu_\perp$ (Fig. \ref{fig:orderparam}), through Eqn. \ref{peak}.  The corresponding limits based on different possible form factors are listed in Table II.

\section{\label{sec:formfactor} Form factor}

The electronic configuration associated with magnetism in URu$_2$Si$_2$ is not known.  Two possibilities are U$^{4+}$ ($5f^2$) or U$^{3+}$ ($5f^3$). \cite{flint2012}  The magnetic form factor, $f_M(|q|)$, which enters into Eqn. \ref{eqn:mag}, is slightly different for the two configurations.  The form factor, which accounts for the spatial distribution of unpaired spin density in the atomic orbitals, can be approximated by the first two terms in a  harmonic expansion,\cite{brown2006magnetic}

\begin{multline}
f_{M}(|q|) =  \langle j_0(|q|) \rangle + \bigg(1-\frac{2}{g_L}\bigg) \langle j_2(|q|) \rangle
\label{eqn:formfactor}
\end{multline}
where $g_L$ is the Land\'e $g$-factor and,
\begin{equation}
\label{eqn:javg}
 \langle j_l (k) \rangle = \int_0^\infty U^2(r) j_l(kr) 4\pi r^2 dr.
 \end{equation}
  Here $U(r)$ is the radial wave function for the unpaired spins and $j_l$ is the $l^{th}$ spherical Bessel function.  
  
The form factors were calculated at the relevant wave vectors using data from Ref. \onlinecite{brown2006magnetic} and are tabulated in Table \ref{tab:magneticdata} for both $5f^2$ and $5f^3$ configurations. 
We choose the $5f^2$ configuration for determining $\mu_\parallel$, consistent with Ref. \onlinecite{broholm1991magnetic}.  For the upper limit on $\mu_\perp$,  we choose the form factor for the $5f^3$ configuration since it produces a more relaxed upper limit.

\section{\label{sec:multscatt} Multiple Scattering}

\begin{figure}[htbp]  
\centering
\includegraphics[width=8cm]{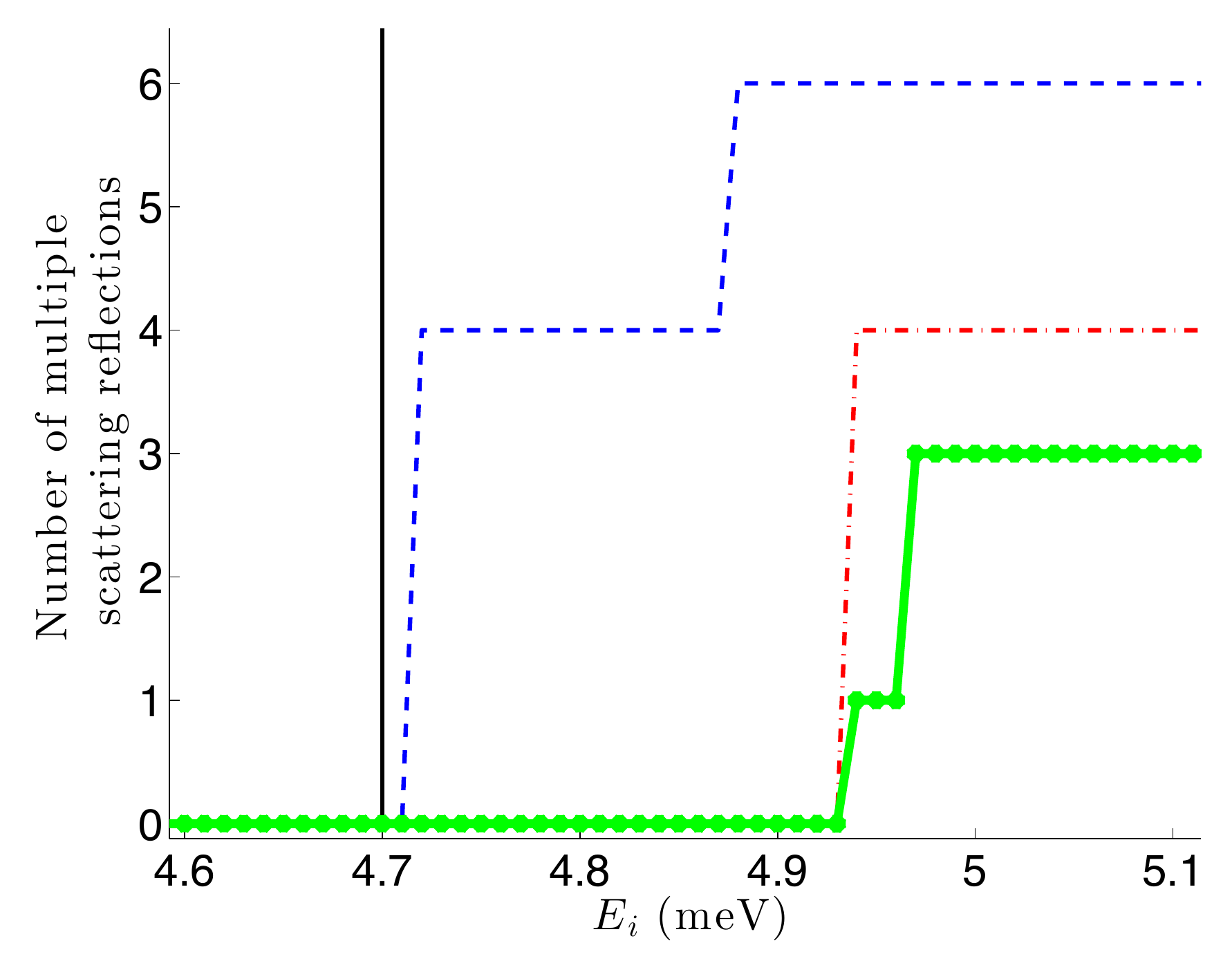}
\caption{ The number of reflections accessible by multiple scattering when the scattering geometry is set to measure (001) (blue dashed line), (003) (red dot-dashed line), and (101) (green line with symbols) for different incident energies ( $E_i$), with ${\bf k}_i$ and ${\bf k}_f$ spanning the $(H0L)$ scattering plane.  At $E_i$ = 4.7 meV (black solid line), no multiple scattering is possible for these configurations within a resolution tolerance of 0.05 \AA.}
\label{fig:multscatt}
\end{figure}

Multiple elastic scattering can occur when a sphere of radius $|\bf{k}_i|$ centered at $\bf{k}_i$ passes through reciprocal lattice points other than the origin and the desired reflection.   Figure \ref{fig:multscatt} shows how many such reflections (with distances within 0.05 \AA$^{-1}$) exist for URu$_2$Si$_2$ arranged in the $(H0L)$ scattering plane, as a function of incident energy.  Our chosen incident energy of 4.7 meV is free of multiple scattering (shown by the black line).

\end{document}